\documentclass[a4paper,aps,onecolumn,showpacs,showkeys,nofootinbib,preprintnumbers,superscriptaddress,amsmath,amssymb,amsfonts]{revtex4}
\usepackage{graphicx}
\usepackage{bm,natbib}
\usepackage{amsmath}
\usepackage[english]{babel}
\usepackage[T1]{fontenc}
\usepackage{amssymb}
\usepackage{amsfonts}
\usepackage{epsfig}
\usepackage{colordvi}
\usepackage{psfrag}
\usepackage{color}

\newcommand{\Lagr}{\mathcal{L}}
\newcommand{\G}{\mathcal{G}}

\usepackage{dcolumn}
\usepackage{multirow}
\usepackage{hyperref}
\usepackage{epstopdf}
\usepackage{bm}
\usepackage{times}
\usepackage{slashed}

\hypersetup{
  colorlinks=true,        
  linkcolor=blue,         
  citecolor=magenta,      
}

\begin{document}
\title{Energy Conditions in Gauss-Bonnet Gravity}


\author{Francesco Bajardi}
\email{f.bajardi@ssmeridionale.it}
\affiliation{Scuola Superiore Meridionale, Largo San Marcellino 10, I-80138, Naples, Italy.}
\affiliation{INFN Sez. di Napoli, Compl. Univ. di Monte S. Angelo, Edificio G, Via Cinthia, I-80126, Naples, Italy.}

\date{\today}

\begin{abstract}
We study the Energy Conditions in modified $f(\G)$ gravity, with $\G$ being the topological Gauss-Bonnet term. Then we use the cosmographic parameters to constrain the functional form of the gravitational action and investigate the possibility to have standard inflation in the early time. Specifically, we select models containing symmetries within the modified $f(\G)$ theory and obtain conditions for which i) the energy conditions can be violated and ii) the magnitude of the slow-roll parameters is small, thus suggesting that under given limits the analyzed theory can potentially trace the cosmic history both at early and at the late times. 
\end{abstract}

\pacs{---}


\maketitle

\section{Introduction}
The success of General Relativity (GR) as the best accepted theory of gravity is undoubted, especially after the recent gravitational waves detection \cite{Abbott:2016blz} and black hole observation \cite{Akiyama:2019fyp}. Providing a new interpretation for the gravitational interaction, GR obtained several successes throughout the years, from the solar system up to cosmological scales. It was the first theory to successfully predict the precession of the perihelion of Mercury in its orbit around the sun, the light deflection, the cosmological epochs crossed by the universe evolution, \emph{etc}. However, in spite of this wide success and the large amount of predictions, GR also suffers from several shortcomings at any scale of energy. Disagreements with observations mostly occur at galactic and cosmological scales. Regarding the former, theoretical predictions on the galaxy rotation curve are in contrast with experiments, as the speed of the farthest stars orbiting around a given galaxy is experimentally higher than theoretically expected \cite{Bosma:1981zz}. This disagreement is so far addressed to an unclustered form of matter called Dark Matter. Similarly, in the cosmological context, the today observed acceleration of the universe is addressed to the so called Dark Energy. Dark Matter and Dark Energy are supposed to account for the major part of the universe content, though they have never been detected directly. In the small--scale regime, similar considerations apply, since GR formalism cannot be merged with that of the other interactions, which in turn are Yang-Mills gauge theories \cite{Horava:2009uw, DeWitt:1967ub, Engle:2007uq, Alexandrov:2002br, Maldacena:1997re, Nilles:1983ge}. This yields the impossibility of dealing with a grand unified theory, including all the fundamental interactions. On the one hand, the two-loop effective action of GR exhibits incurable UV divergences that cannot be renormalized by means of standard renormalization techniques \cite{Goroff:1985th}. On the other hand, any attempt to quantize the theory and construct a self consistent quantum formalism has failed. In light of these shortcomings, several GR modifications began to be considered. Among the alternatives, special interest was gained by those theories extending the Einstein-Hilbert action to functions of second-order curvature invariants (for reviews on the topic see \cite{Sotiriou:2008rp, DeFelice:2010aj, Clifton:2011jh, Nojiri:2006gh, Capozziello:2019klx} and reference therein). 

Extending GR, in general, can allow to explain the observed acceleration of the Universe as a geometric contribution, without need to invoke the presence of "dark" sectors \cite{Capozziello:2011et, Setare:2012ry, Nojiri:2006ri, Bajardi:2020xfj}. This is due to the fact that extended actions lead to higher-order field equations that can be usually recast as
\begin{equation}
F G_{\mu \nu} = T_{\mu \nu}^{(curv)} + T_{\mu \nu}^{(Matter)},
\label{genfield}
\end{equation}
where $F$ is a generic function depending on the theory considered, $G_{\mu \nu}$ the Einstein tensor $G_{\mu \nu} = R_{\mu \nu} - 1/2 g_{\mu \nu} R$ (with $R_{\mu \nu}$ and $R$ being the Ricci tensor and the Ricci scalar, respectively), $T_{\mu \nu}^{(Matter)}$ the energy-momentum tensor of matter fields, and $T_{\mu \nu}^{(curv)}$ the so called Energy-Momentum tensor of the gravitational field. The latter includes extra geometric terms that, in principle, can mimic the role of Dark Energy. A straightforward example is given by the $f(R)$ extension of GR, whose action reads as:
\begin{equation}
S = \int \sqrt{-g} f(R) \, d^4x,
\end{equation}
with $f(R)$ being a general function of the scalar curvature $R$. The variation of the action with respect to the metric tensor, yields in vacuum the following fourth-order field equations:
\begin{equation}
f_R G_{\mu \nu} = \frac{1}{2} g_{\mu \nu} \left[f(R) - R f_R(R) \right] + \nabla_\mu \nabla_\nu f_R(R) - g_{\mu \nu} \Box f_R(R).
\label{f(R)FE}
\end{equation}
In the above equation, $f_R$ is the first derivative of $f(R)$ with respect to $R$, $\nabla_\mu$ is the covariant derivative and $\Box \equiv g^{\mu \nu} \nabla_\mu \nabla_\nu$ the D'Alembert operator. By comparing Eq. \eqref{f(R)FE} with Eq. \eqref{genfield} we notice that $F= f_R$ and 
\begin{equation}
    T_{\mu \nu}^{(curv)} = \frac{1}{2} g_{\mu \nu} \left[f(R) - R f_R(R) \right] + \nabla_\mu \nabla_\nu f_R(R) - g_{\mu \nu} \Box f_R(R).
\end{equation}
For a complete discussion regarding $f(R)$ theories see \cite{Capozziello:2009nq, Nojiri:2017qvx}. Besides $f(R)$ gravity, one can also include other higher-order curvature invariants into the action, as standard in the literature, such as $R^{\mu \nu} R_{\mu \nu}$ \cite{Gron:2002rg, Alty:1994xj}, the Kretschmann scalar $R^{\mu \nu p \sigma} R_{\mu \nu p \sigma}$ \cite{Cherubini:2002gen, Stelle:1976gc}, $\Box R$ \cite{Wands:1993uu, Gottlober:1989ww}, \emph{etc.}

In this paper we consider an extension of the Einstein-Hilbert action, including a function of the so called Gauss-Bonnet term $\G$, defined through a linear combination of the Riemann tensor square $R^{\mu \nu \rho \sigma} R_{\mu \nu \rho \sigma}$, the Ricci tensor square $R^{\mu \nu}R_{\mu \nu}$ and the Ricci scalar square $R^2$. More precisely, it reads:
\begin{equation}
\G = R^2 - 4 R^{\mu \nu}R_{\mu \nu} +R^{\mu \nu \rho \sigma} R_{\mu \nu \rho \sigma}.
\label{GBDef}
\end{equation}
From the above definition, one can show that in 3+1 dimensions $\G$ reduces to a boundary term, while in 2+1 dimensions (or less) it vanishes identically. As a consequence, the action $S = \int \sqrt{-g} (\G + R) d^4x$ is completely equivalent to the Einstein-Hilbert one at the level of equations, since $\G$ can be integrated out. Nevertheless, the Gauss-Bonnet scalar is often considered in higher-dimensional theories, such as Chern-Simons \cite{Gomez:2011zzd, Aviles:2016hnm, Bajardi:2021hya}, Born Infield \cite{Hao:2003ib} or Lovelock \cite{Cvetkovic:2016ios, Deruelle:1989fj, Myers:1988ze} gravity, where it naturally emerges as a non-trivial contribution. To avoid the problem of triviality in four dimensions, in recent works four-dimensional Gauss--Bonnet gravity has been considered by multiplying $\G$ by a dimension-depending coupling constant, which diverges when $D=4$~\cite{Glavan:2019inb, Gurses:2020rxb}.

Another way to deal with the Gauss--Bonnet scalar in four dimensions, is to consider a function of $\G$, which starts to be trivial in three dimensions. The advantages of the latter approach is twofold. On the one hand, as $\G$ naturally arises in gauge theories of gravity, it can result helpful in addressing issues related to renormalizability and adaptation to Quantum Field Theory. On the other hand, the function $f(\G)$ carries extra terms in the field equations which can explain the late-time evolution of the Universe as a geometric contribution, similarly to $f(R)$ gravity.

The $f(\G)$ model, with a starting action reading as
\begin{equation}
    S = \frac{1}{2}\int \sqrt{-g} \left[R + f(\G) \right] d^4 x,
    \label{actionr+f}
\end{equation}
has been widely studied in the literature (see \emph{e.g.} \cite{Capozziello:2014ioa, Nojiri:2005jg, Uddin:2009wp, DeFelice:2009rw, Davis:2007ta}), due to the capability of $f(\G)$ to play the role of a dynamical cosmological constant provided by geometry, preserving GR as $f(\G)$ vanishes. 

Cosmological and astrophysical applications of this theory are analyzed \emph{e.g.} in \cite{DeFelice:2008wz, Zhong:2018tqn, S.Silva:2018irj}. In many works, Eq. \eqref{actionr+f} is further generalized by considering a function of both $R$ and $\G$ into the action \cite{Benetti:2018zhv, Capozziello:2016eaz, Shamir:2017ndy}. 

Another possibility is to evaluate an action only containing the function $f(\G)$, without imposing the GR limit as a requirement, namely
\begin{equation}
       S = \int \sqrt{-g}  f(\G) \, d^4 x.
       \label{actionfg}
\end{equation}
Though GR cannot be exactly recovered, it is still possible to get the same results under some approximations, as pointed out in Refs. \cite{Bajardi:2020osh, Bajardi:2019zzs}.

Here we start from the actions \eqref{actionr+f} and \eqref{actionfg}, and select the functions by means of the so called Noether symmetry approach, a selection criterion aimed at finding symmetries of the point-like Lagrangian related to a gravitational theory \cite{Bajardi:2022ypn}. Then, we further constrain these selected models by means of the Energy Conditions (ECs) and the cosmographic parameters.

More precisely, this paper is organized as follows. In Sec. \ref{ECETG} we overview the main aspects of the ECs and the roles they play in gravitational theories, such as extended theories of gravity. In Sec. \ref{sec:noeth} we introduce the Noether symmetry approach and apply the latter to two modified Gauss--Bonnet models, to select the functional form of the action. In Secs. \ref{ECCOS} and \ref{ECCOS1} we study the ECs violation in the previously selected $f(\G)$ and $R+ f(\G)$ gravitiies, respectively, also investigating the slow-roll inflation. Finally, in Sec. \ref{concl} we conclude the work outlining the results and future perspectives. 

\section{Energy Conditions in Extended Gravity}\label{ECETG}
As it is well known, in GR the ECs only rely on the Energy-Momentum tensor of matter, which automatically satisfies all the four inequalities on the pressure and the energy density. The ECs are also required as a support for several physical foundations, such as the no hair theorem \cite{Sultana:2018fkw} or the laws of black hole thermodynamics \cite{Nojiri:2004pf}. Specifically, the ECs (in natural units) read as:
\begin{equation}
\begin{split}
&\text{Null Energy Condition (NEC)} \to \rho + p \ge 0
\\
& \text{Weak Energy Condition (WEC)} \to \rho \ge 0 \; ; \; \rho + p \ge 0
\\
&\text{Dominant Energy Condition (DEC)} \to \rho - |p| \ge 0
\\
&\text{Strong Energy Condition (SEC)} \to \rho + p \ge 0 \; ; \; \rho + 3p \ge 0 \;,
\end{split}
\end{equation}
where $\rho$ is the energy density and $p$ the pressure. The last relation implies that gravity must be attractive, while the others require the pressure and the energy to be non-negative. In GR, where the only energy momentum tensor is that of the standard matter, all the ECs are identically satisfied. 

Modifications of the gravitational action make the field equations more complex and can introduce branching behavior in the solution space. This is the case with Gauss-Bonnet gravity, as pointed out in Refs. \cite{Henneaux:1987zz, Deruelle:1989fj}. Therefore, EC violation may imply the presence of a branched Hamiltonian, depending on the form of the selected function.

A branched or multivalued Hamiltonian typically arises in systems where the energy landscape or phase space exhibits multiple sheets or branches. Such structures appear in classical and quantum systems with constraints or symmetry-breaking phenomena. These Hamiltonians often signify a system that can switch between different modes or configurations, leading to multivalued dynamics depending on the initial conditions or energy levels.

Modified models, such as $f(R,\G)$ gravity, often yield multiple solutions under the same boundary or initial conditions, depending on the form of the function \cite{DeFelice:2010aj}. In this context, branches can allow for the modeling of diverse phenomena, \emph{e.g.} $f(\G)$-based inflationary solutions and late-time dark energy behaviors, though some of them may correspond to unstable solutions \cite{delaCruz-Dombriz:2011oii}.

In fact, as discussed in \cite{DeFelice:2009ak}, theories based on $f(R,\G)$ may encounter the problem of ghost modes, even within the context of cosmological perturbations, where superluminal modes can arise. In particular, this issue is critical for $f(R,\G)$ gravity because the governing equation for the evolution of primordial curvature perturbations inherently features superluminal modes. This arises primarily due to the Gauss-Bonnet invariant introducing higher-order derivatives relative to the metric. Consequently, the Lagrangian cannot be reformulated in a canonical form, leading to a Hamiltonian that exhibits linear instability.

This instability originates from a mismatch between the parameters required to define the problem and the actual dependencies of the Lagrangian. Even with an appropriate gauge choice, higher-order metric derivatives persist in the field equations of $f(R,\G)$ gravity, potentially giving rise to ghost modes. This issue has been addressed in studies such as \cite{Astashenok:2015haa, Nojiri:2018ouv}, where the Lagrange multiplier formalism is employed to mitigate ghost instabilities. This approach effectively removes ghost degrees of freedom in both $f(\G)$ and $f(R,\G)$ gravity models, enabling these frameworks to produce ghost-free primordial curvature perturbations.

To establish ghost-free conditions, the authors in these references apply a conformal transformation to the metric and work within the Jordan frame. They couple the Gauss-Bonnet invariant with a scalar field and propose an \emph{ad hoc} potential form. Subsequently, by expanding the metric around flat spacetime and analyzing the resulting field equations, it is shown that, in the Jordan frame, the equations only involve first- and second-order derivatives of $\delta g_{\mu \nu}$  with respect to spacetime coordinates. However, the scalar field lacks dynamical behavior, ensuring that no additional degrees of freedom arise compared to GR.

Summarizing, it is not clear whether a general connection between ECs and branched Hamiltonians can be established (especially because not all the $f(R,\G)$ models violate the energy conditions), apart from the fact that both features arise due to higher-order field equations and the non-canonicity of the Lagrangian. Therefore, although there is no evident direct link between the two, the occurrence of the former may imply the appearance of the latter, and \emph{vice versa}. A more comprehensive study of the Hamiltonian structure in higher-order modified gravity is needed to investigate the possibility of a direct link between these two features.

From a cosmological point of view, the ECs can be used along with the cosmographic parameters to test the validity of the theory, as well as to constrain the starting gravitational actions leading to the late-time cosmic expansion. This approach is straightforwardly discussed \emph{e.g.} in Refs. \cite{Capozziello:2011hj, Capozziello:2014bqa, Santos:2007bs, Garcia:2010xz}, where different alternatives to GR are considered. 

Following a similar procedure, in Sec. \ref{ECCOS} we apply the ECs to the action \eqref{actionfg}, showing that the Einstein tensor can be isolated even when the GR limit is not assumed as a requirement. Then, in Sec. \ref{ECCOS1}, we compare the result with that provided by $R + f(\G)$ gravity. In both cases, we consider functions selected by Noether symmetries and show that the effective energy density and the effective pressure can be written in terms of the geometry. As a result, in a Friedmann-Lemaitre-Robertson-Walker space-time, the cosmographic parameters along with the requirements for ECs violation allow to further constrain these models. 
Moreover, the effective energy density and pressure of modified Gauss-Bonnet gravity can be also used in order to find the explicit form of the slow--roll parameters and to check whether these models admit a cosmological inflation in the early time. 

Inflationary model was introduced by A. Linde and A. Guth \cite{Linde:1981mu, Guth:1980zm} to address evidences provided by the cosmological data at the early time. Inflation is usually thought as generated by a scalar field $\phi$, called inflaton, which is supposed to be the responsible for the accelerated expansion of the universe. Soon after, the Starobinsky model \cite{Starobinsky:1982ee} showed that the additional geometric contributions occurring in modified theories of gravity can be intended as an effective scalar potential capable of driving the inflation. Therefore, nowadays inflation can be realized in several ways \cite{DeLaurentis:2015fea, Albrecht:1982wi, Freese:1990rb, Polarski:1992dq, Linde:1993cn}.

In Sec. \ref{SRinfl} we consider the so called \emph{new inflation} or \emph{slow--roll inflation}, according to which inflation is driven by a scalar field rolling down a potential energy hill. Inflation occurs as soon as the scalar field rolling is slow with respect to the Universe expansion. 

By using the ECs, we require the slow--roll parameters to be small during inflation. This allows to put some constraints on the $f(\G)$ function, selecting those compatible with inflationary universes. 
\section{Selecting cosmological models via Noether symmetries}
\label{sec:noeth}
In this section we provide a short overview of the Noether Symmetry approach and apply the latter to modified gravity models containing the Gauss--Bonnet term, in a cosmological and spatially-flat background. The research for symmetries in this context gives a selection criterion for viable physical models based on theoretical arguments and can be helpful in several ways. On the one hand, the presence of symmetries can allow to decrease the complexity of the modified Friedmann equations, by reducing the dynamics and allowing to find analytic solutions. On the other hand, the method offers a valid approach to select the functional form of the action, to be tested by cosmological observations. 

Let us briefly revisit the key characteristics of the Noether symmetry approach, illustrating the technique employed to identify symmetries within the gravity theories mentioned earlier. We utilized this approach to choose the functional structure of gravitational actions, streamline dynamics, and ultimately discover exact solutions.

A general transformation involving time $t$ and coordinates $q^i$, can be written in terms of the infinitesimal generators $\xi(t, q^i)$ and $\eta^i(t, q^i)$ as
\begin{equation}
\left\{\begin{array}{l}\tilde{t}=t+\epsilon \xi +O\left(\epsilon^{2}\right) \\ 
\tilde{q}^{i}=q^{i}+\epsilon \eta^{i}+O\left(\epsilon^{2}\right)\end{array}\right.  \,,
\label{trans}
\end{equation}
providing the following generator
\begin{equation}
\mathcal{X} =  \xi \frac{\partial }{\partial t} + \eta^i \frac{\partial }{\partial q^i} \;.
\label{generator}
\end{equation} 
Noether's theorem affirms that the generator \eqref{generator} is a symmetry for a given Lagrangian $\Lagr$, iff
\begin{equation}
X^{[1]} \Lagr + \dot{\xi} \Lagr = \dot{g}(t,q^i) \;,
\label{Teorema}
\end{equation}
with $X^{[1]}$ being defined as:
\begin{equation}
X^{[1]} =  \xi \frac{\partial }{\partial t} + \eta^i \frac{\partial }{\partial q^i} + (\dot{\eta}^i - \dot{q}^i \dot{\xi}) \frac{\partial}{\partial \dot{q}^i} \;
\label{firstprolong}
\end{equation} 
and with $g$ being a generic function of the variables of the considered space of configurations ${\cal Q}\equiv\{q_i\}$, namely the minisuperspace, which is a restriction of the infinite-dimensional superspace of the ADM model \cite{Arnowitt:1962hi}. Moreover, provided that the condition in Eq.~\eqref{Teorema} holds, it can be shown that the following quantity is a first integral of motion:
\begin{equation}
I(t,q^i,\dot{q}^i) = 	\displaystyle \xi \left(\dot{q}^i \frac{\partial \Lagr}{\partial \dot{q}^i} - \Lagr \right) - \eta^i \frac{\partial \Lagr}{\partial \dot{q}^i} + g(t,q^i) \;.
\end{equation}
It is worth stressing the importance of the presence of the above the conserved quantity since, if the latter exists, then a coordinate system containing new cyclic variables can be properly chosen. Further details on Noether's approach and applications to cosmology and astrophysics in different backgrounds can be found \emph{e.g.} in \cite{Bajardi:2022ypn}
\subsection{The case of $R+f(\G)$ gravity}
The first application of the Noether's approach is to $R+ f(\G)$ gravity; to this purpose, let us start by evaluating the following gravitational action:
\begin{equation}
S = \int \sqrt{-g} \left[\frac{R}{2} + f(\G) \right] d^4 x,
\label{actionR+f}
\end{equation}
in a cosmological and spatially flat space-time of the form
\begin{equation}
ds^2 = dt^2 - a(t)^2 d\textbf{x}^2,
\label{linelement}
\end{equation}
with $a(t)$ being the scale factor, $R$ the ricci scalar and $\G$ the Gauss-Bonnet topological surface term defined in Eq. \eqref{GBDef}. In order to search for symmetries and apply the condition \eqref{Teorema}, we must first find the point-like Lagrangian by using the Lagrange multipliers method. To this purpose, notice that we can either assume $R$ and $a$ as independent field, or replace the cosmological expression of the Ricci curvature into the action. The latter approach results more straightforward, as it allows to reduce the minisuperspace to two dimensions and obtain a suitable cosmological Lagrangian. However, both scenarios are equivalently valid and lead to the same solutions from a symmetry point of view. By pursuing this procedure, the Ricci scalar is replaced by its cosmological form and the only dynamical degrees of freedom turn out to be the Gauss-Bonnet term and the scale factor. Adopting the Lagrange multipliers scheme, Eq. \eqref{actionR+f} can be recast as:
\begin{equation}
    S = \int \sqrt{-g} \left[\frac{R}{2} + f(\G) - \lambda \left( \G - \tilde{\G} \right) \right] d^4x,
\end{equation}
where $\tilde{\G}$ is the Gauss--Bonnet scalar written in terms of the given background. Choosing the space-time in Eq. \eqref{linelement}, the above action takes the form
\begin{equation}
    S = \int a^3 \left[-6 \left(\frac{\ddot{a}}{a} + \frac{\dot{a}^2}{a^2} \right) + f(\G) - \lambda \left( \G - 24 \frac{\dot{a}^2 \ddot{a}}{a^3} \right) \right] dt.
    \label{actioncosmo}
\end{equation}
In the above expression, we have integrated the three-timensional hypersurface and considered the cosmological expression of the Ricci scalar and the Gauss--Bonnet term, with respect to the line element \eqref{linelement}, \emph{i.e.}
\begin{equation}
    R = -6 \left(\frac{\ddot{a}}{a} + \frac{\dot{a}^2}{a^2} \right), \qquad \G = 24 \frac{\dot{a}^2 \ddot{a}}{a^3}.
    \label{cosmoexprGR}
\end{equation}
By varying Eq. \eqref{actioncosmo} and using the least action principle, it is possible to get $\lambda = a^3 f_\G$, with $f_\G$ denoting the first derivative of $f(\G)$ with respect to $\G$. After doing so and integrating out second derivatives, we finally find the Lagrangian, which reads as:
\begin{equation}
\Lagr = 3 a \dot{a}^2 + a^3(f- \G f_\G) - 8 \dot{a}^3 \dot{\G} f_{\G \G}.
\label{lagr R+f(G)}
\end{equation}
It is important to emphasize that the minisuperspace reduction is applicable solely due to the linear presence of the scalar curvature in the action. Without this linearity, the integration of higher derivatives becomes nontrivial, making the derivation of a point-like Lagrangian impractical. Following our scheme, the minisuperspace turns out to be independent of the curvature scalar, so that the minisuperspace is $\mathcal{Q} = \{a,\G\}$ and the symmetry generator can be recast as:
\begin{equation}
\mathcal{X} = \xi \partial_t + \alpha \partial_a + \gamma \partial_\G,
\label{GENR+FG}
\end{equation}
with $\alpha \equiv \alpha(t,a,\G)$ and $\gamma \equiv \gamma(t,a,\G)$ being the only non-vanishing components of the infinitesimal generator $\eta^i$. Considering now the first prolongation Noether's vector in Eq. \eqref{firstprolong} and the Lagrangian \eqref{lagr R+f(G)}, the identity \eqref{Teorema} provides the system
\begin{equation}
\begin{cases}
\displaystyle \frac{\alpha}{a} + 2 \partial_a \alpha - \partial_t \xi = 0
\\
\displaystyle \gamma f_{\G \G \G} + f_{\G\G} \left(3 \partial_a \alpha + \partial_\G \gamma - 3 \partial_t \xi \right) = 0
\\
\displaystyle (f - \G f_{\G})(3 \alpha + a \partial_t \xi) - \gamma a \G f_{\G\G} = 0
\\
\xi=\xi(t), \,\,\,\, \alpha = \alpha(a), \,\,\,\,\, \gamma = \gamma(\G),
\end{cases}
\end{equation}
whose only solution is:
\begin{equation}
\begin{cases}
{\cal{X}} = t \partial_t  + \frac{1}{3} a \partial_a - 4 \G \partial_\G
\\
f(R,\G) = \displaystyle \frac{R}{2} + f_0 \sqrt{\G} + f_1 \G,
\end{cases}
\label{symmetryfunc1}
\end{equation}
meaning that $\xi$ is linearly dependent on the cosmic time and $\alpha = \frac{1}{3} a$, $\gamma = - 4 \G$. The former condition occurs whenever the starting lagrangian is canonical, as shown in Ref. \cite{Bajardi:2023byv}. The application of Noether's approach eventually manifests that the only function containing extra symmetries is $f(\G) = f_0 \sqrt{\G} + f_1 \G$. However, the linear contribution of the Gauss--Bonnet scalar turns out to be trivial due to the topological nature of $\G$. The other term, that is $f_0 \sqrt{\G}$, non-trivially contributes to the dynamics and leads to time power-law solutions for the scale factor, as shown in Sec. \ref{ECCOS1}. Also notice that, under given cosmological contexts, $\G$ is simply proportional to $R^2$, as discussed in \cite{Bajardi:2020osh}, implying that $\sqrt{\G}$ dynamically behaves like the scalar curvature. In this framework, it is interesting to notice that the only action selected by Noether symmetries is the one mimicking the Einsten-Hilbert's.
\subsection{The case of $f(\G)$ gravity}
When the Ricci curvature scalar is not considered as a part of the starting action, namely when the latter becomes
\begin{equation}
S = \int \sqrt{-g}  f(\G) \, d^4 x,
\label{actionf}
\end{equation}
the Lagrange multipliers method, in the background \eqref{linelement} and with constraint \eqref{cosmoexprGR}, provides
\begin{equation}
S = \int \left[a^{3} f(\mathcal{G}) - \lambda \left\{ \mathcal{G} - 24 \frac{\dot{a}^2 \ddot{a}}{a^3} \right\} + \Lagr_m \right] d^{4}x \;,
\label{azione con lambda}
\end{equation}
where $\Lagr_m$ is the matter Lagrangian. The vatiation of Eq. \eqref{azione con lambda} with respect to $\mathcal{G}$, allows to find
\begin{equation}
\delta S = \frac{\partial S}{\partial \mathcal{G}} \delta \mathcal{G} = a^{3} f_{\G}(\mathcal{G}) - \lambda = 0, \;\;\;\;\;\; \lambda = a^{3} f_{\G}(\mathcal{G}) \;.
\end{equation}
Substituting the outcome into Eq. \eqref{azione con lambda} and eliminating the second derivative through integration, the Lagrangian eventually assumes the following structure:
\begin{equation}
\Lagr =  a^3 [f(\G) - \G f_{\G}(\G)] - 8 \dot{a}^3 \dot{\G} f_{\G\G}(\G) ,
\label{lagrangiana nD}
\end{equation}
where we the matter Lagrangian contribution has been neglected for simplicity. Now we adopt the Noether Symmetry existence condition \eqref{Teorema}, assuming that the generator
\begin{equation}
\mathcal{X} = \xi(a,\G,t) \partial_t + \alpha(a,\G,t) \partial_a + \beta(a,\G,t) \partial_\G\,
\end{equation} 
is a symmetry for the Lagrangian \eqref{lagrangiana nD}. The approach yields a system of two differential equations, completed with the constraints on the infinitesimal generators $\alpha, \beta, \xi$:
\begin{equation}
\begin{cases}
\displaystyle 3 \, a^2 \alpha (f- \G f_{\G}) - 24 \dot{a}^2 \dot{\G} f_{\G \G}\partial_t \alpha - 
 a^3 \left[\beta f_{\G \G}' - (f - \G f_{\G}) \partial_t \xi \right] = 0
\\
\displaystyle   \beta f_{\G \G \G} + f_{\G \G} \left( 3 \partial_a \alpha + \partial_\G \beta - 3 \partial_t \xi \right) = 0
\label{System}
\end{cases}
\end{equation}
with $ \xi = \xi(t)\,,  \alpha = \alpha(a,t)\,, \beta = \beta(t, \G)\,,  g = 0\,.$ Here, we discard \emph{a priori} the possibility $f_{\G \G} = 0$, as it would only lead to non-physical results. From the resolution of the system, it turns out that the only non-trivial symmetry is given by
\begin{equation}
\alpha = \alpha_0 a\,, \quad \beta = - 4 \xi_0 \G\,, \quad \xi = \xi_0 t + \xi_1\,, \,\, f(\G) = \frac{ 4 f_0 \xi_0}{3\alpha_0  + \xi_0} \G^{\frac{3 \alpha_0  + \xi_0}{4 \xi_0}}, 
\label{soluznoeth}
\end{equation}
with the obvious constraint $\alpha_0 \neq  \xi_0$. The resulting action, thus, can be written in a more compact form, by defining 
\begin{equation}
\frac{3 \alpha_0 + \xi_0}{4 \xi_0} = k
\end{equation}
and incorporating the coefficient of $\G^k$ into $f_0$. In this way, within all possible actions of the form \eqref{actionf}, the only one containing symmetries is
\begin{equation}
    S = \int \sqrt{-g} \, \G^k \, d^4 x.
\end{equation}
As mentioned before, when $k=1/2$ the above action is capable of mimicking GR in some cosmological space-times. In the analysis of the next sections, we aim to constrain the value of $k$ by means of ECs and slow-roll inflation.
\section{Energy Conditions in $f(\G)$ Cosmology} \label{ECCOS}
Once we introduced the main aspects of the ECs in extended theories of gravity and selected the cosmological modified Gauss--Bonnet models based on symmetry considerations, we now write the ECs for $f(\G) \sim \G^k$ model, constraining the latter by cosmographic parameters. Let us begin by considering the $f(\G)$ action
\begin{equation}
S = \int \sqrt{-g} \; f(\G) \; d^4 x  + S^{(m)} \;,
\label{azione}
\end{equation}
with $S^{(m)}$ being the matter action. The variation with respect to the metric yields the following field equations \cite{Bajardi:2020osh, Nojiri:2005jg, Bamba:2017cjr}:
\begin{eqnarray}
&&T_{\mu \nu} + \frac{1}{2} g_{\mu \nu} f(\G) - \left(2R R_{\mu \nu} - 4 R_{\mu p} R^p_{\,\,\,\nu} + 2 R_\mu^{\,\,\, p \sigma \tau} R_{\nu p \sigma \tau} - 4 R^{\alpha \beta} R_{\mu \alpha \nu \beta}\right) f_\G(\G) +\nonumber
\\
&& + \left(2R \nabla_\mu \nabla_\nu +4 G_{\mu \nu} \Box - 4 R_{\{ \nu}^p \nabla_{\mu \}}\nabla_p + 4 g_{\mu \nu} R^{p \sigma} \nabla_p \nabla_\sigma - 4 R_{\mu \alpha \nu \beta} \nabla^\alpha \nabla^\beta \right) f_\G(\G) = 0 \;,
\end{eqnarray}
where $T_{\mu \nu}$ represents the matter energy-momentum tensor coming from the variation of $S^{(m)}$ with respect to $g^{\mu \nu}$. Isolating the Einstein tensor $G_{\mu \nu}$ in all terms in which it appears, Eq. \eqref{field eq} can be recast as:
\begin{eqnarray}
 G_{\mu \nu} (2 R -4 \Box ) f_\G(\G) &=&  T_{\mu \nu} - \Big[ \left(R^2 - 4 R_{\mu p} R^p_{\,\,\,\nu} + 2 R_\mu^{\,\,\, p \sigma \tau} R_{\nu p \sigma \tau} - 4 R^{\alpha \beta} R_{\mu \alpha \nu \beta}\right) f_\G(\G)  + \nonumber
\\
&-& \left(2R \nabla_\mu \nabla_\nu - 4 R_{\{ \nu}^p \nabla_{\mu \}}\nabla_p + 4 g_{\mu \nu} R^{p \sigma} \nabla_p \nabla_\sigma - 4 R_{\mu \alpha \nu \beta} \nabla^\alpha \nabla^\beta \right) f_\G(\G)  - \frac{1}{2} g_{\mu \nu} f(\G) \Big] \;.
\label{field eq}
\end{eqnarray}
Notice that the term in the square bracket of the RHS can be intended as the effective Energy-Momentum tensor of the gravitational field, introduced in Eq. \eqref{genfield}, while $(2 R -4 \Box ) f_\G(\G)$ accounts for the function $F$. Therefore, the components of the Einstein tensor can be interpreted as the analogue of energy density and pressure, namely 
\begin{equation}
G_0^0 = \frac{1}{(2 R -4 \Box ) f_\G(\G)}\left( \rho_{\G} + \rho_0 \right) \qquad  G^i_j = - \frac{\delta^i_j }{(2 R -4 \Box ) f_\G(\G)}\left( p_{\G} + p_0\right),
\end{equation}
where $\rho_0$ and $p_0$ are matter density and pressure, respectively, while $\rho_\G$ and $p_\G$ are the effective energy density and pressure provided by extra geometric terms in $f(\G)$ gravity. In a spatially flat cosmological universe of the form \eqref{linelement}, the field equations of $f(\G) = f_0 \G^k$ gravity can be written as:
\begin{eqnarray}
 G_{0}^0 = \frac{1}{(2 R -4 \Box ) f_\G(\G)} && \left\{ \rho_0 + \frac{f_0}{2} \G^{k-3} \left[-24 k (k-1) H^2 \left(\G \ddot{\G}+(k-2) \dot{\G}^2\right) \right. \right. \nonumber 
 \\
 && \left. \left. -72 k \G^2 H^4-48 k \G H^2 \left(2 \G (\dot{H} + H^2) +(k-1) H \dot{\G}\right)+\G^3 \right] \right\} ,
   \label{com00}
\end{eqnarray}

\begin{eqnarray}
G_{1}^1 =-\frac{1}{(2 R -4 \Box ) f_\G(\G)} && \left\{  p_0 + \frac{f_0}{2} \G^{k-3} \left[16 k(k-1) (\dot{H} + H^2) \left(\G \ddot{\G}+(k-2) \dot{\G}^2\right) \right. \right. \nonumber
\\
&&  + 24 k \G^2 H^4+16 k \G (\dot{H} + H^2) \left(3 \G (\dot{H} + H^2)+2 (k-1) H \dot{\G} \right) \nonumber
\\
&&  \left. \left. +24 k \G H^2 \left(4 \G (\dot{H} + H^2) +(k-1) H \dot{G}\right)-\G^3\right] \right\},
   \label{com11}
\end{eqnarray}

with $H$ being the Hubble constant $H \equiv \dot{a}/a$.
The second term in the RHS of Eq. \eqref{com00} is the cosmological density of the gravitational field $\rho_\G$, as the same term in Eq. \eqref{com11} is the pressure. Specifically:
\begin{eqnarray}
\rho_\G = \frac{f_0}{2} \G^{k-3} && \left[-24 k (k-1) H^2 \left(\G \ddot{\G}+(k-2) \dot{\G}^2\right) \right. \nonumber
\\
&& \left. -72 k \G^2 H^4-48 k \G H^2 \left(2 \G (\dot{H} + H^2) +(k-1) H \dot{\G}\right)+\G^3 \right] ,
\label{density}
\end{eqnarray}
\begin{eqnarray}
p_\G = \frac{f_0}{2} \G^{k-3} && \left[16 k(k-1) (\dot{H} + H^2) \left(\G \ddot{\G}+(k-2) \dot{\G}^2\right)  + 24 k \G^2 H^4 \right. \nonumber
\\
&&  +16 k \G (\dot{H} + H^2) \left(3 \G (\dot{H} + H^2)+2 (k-1) H \dot{\G} \right)  \nonumber 
\\
&& \left. +24 k \G H^2 \left(4 \G (\dot{H} + H^2) +(k-1) H \dot{\G}\right)-\G^3\right] ,
\label{pressure}
\end{eqnarray}

so that each EC can be split in two contributions, namely 
\begin{equation}
\begin{split}
&\text{NEC} \to \rho_0 + \rho_\G  + p_0 + p_\G \ge 0
\\
& \text{WEC} \to \rho_0 + \rho_\G \ge 0 \; ; \; \rho_0 + \rho_\G  + p_0 + p_\G \ge 0
\\
&\text{DEC} \to \rho_0 + \rho_\G  - \left|p_0\right|- \left|p_\G\right| \ge 0
\\
&\text{SEC} \to \rho_0 + \rho_\G  + p_0 + p_\G\ge 0 \; ; \; \rho_0 + \rho_\G  + 3 p_0 + 3p_\G  \ge 0 \;.
\end{split}
\label{ECSP0P}
\end{equation}
Starting from the above inequalities, in what follows we focus on a particular subcase, requiring that the standard matter and the gravitational field must respect the ECs separately. Assuming that the ordinary matter automatically satisfies all the ECs, the only constraints capable of providing the validity range of the parameter $k$ are given by the system:
\begin{equation}
\begin{split}
&\text{NEC} \to \rho_\G+ p_\G \ge 0
\\
& \text{WEC} \to \rho_\G \ge 0 \; ; \; \rho_\G + p_\G \ge 0
\\
&\text{DEC} \to \rho_\G - |p_\G| \ge 0
\\
&\text{SEC} \to \rho_\G + p_\G \ge 0 \; ; \; \rho_\G + 3p_\G \ge 0 \;.
\end{split}
\label{ECGF}
\end{equation}
Replacing Eqs. \eqref{density} and \eqref{pressure} into the system \eqref{ECGF}, the ECs yield
\begin{eqnarray}
NEC \to && 4 f_0 k \G^{k-3} \left[(k-1) \G \left(2 \ddot{\G} \dot{H} -H^2 \ddot{\G}+4 H \dot{\G} \dot{H}+H^3 \dot{\G}\right)\right. \nonumber
\\
&& \left. -(k-1)(k-2) \dot{G}^2   \left(H^2-2 \dot{H}\right) +6 \G^2 \dot{H} \left(\dot{H}+2 H^2\right)\right] \ge 0  \qquad \qquad \qquad \qquad 
    \label{EC0}
\end{eqnarray}
\begin{eqnarray}
&& 4 f_0 k \G^{k-3}  \left[(k-1) \G \left(2 \ddot{\G} \dot{H} -H^2 \ddot{\G}+4 H \dot{\G} \dot{H}+H^3 \dot{\G}\right)\right. \nonumber
\\
&& \left.-(k-1)(k-2) \dot{G}^2   \left(H^2-2 \dot{H}\right) +6 \G^2 \dot{H} \left(\dot{H}+2 H^2\right)\right] \ge 0    \nonumber
   \\
 WEC^{\;\;\; \nearrow}_{\;\;\; \searrow}&&\;\;\;\;\;\;\;\;\;\;\;\;\;\;\;\;\;\;\;\;\;\;\;\;\;\;\;\;\;\;\;\;\;\;\;\;\;\;\;\;{\text{and}} \nonumber
   \\
&& \frac{f_0}{2} \G^{k-3} \left[-24 k (k-1) H^2 \left(\G \ddot{\G}+(k-2) \dot{\G}^2\right)  -72 k \G^2 H^4 \right. \nonumber
\\
&& \left. -48 k \G H^2 \left(2 \G (\dot{H} + H^2) +(k-1) H \dot{\G}\right)+\G^3 \right] \ge  0 \qquad \qquad  \qquad \qquad \,\, \qquad 
   \label{EC111}
\end{eqnarray}
\begin{eqnarray}
&& 4 f_0 k \G^{k-3} \left[(k-1) \G \left(2 \ddot{\G} \dot{H} -H^2 \ddot{\G}+4 H \dot{\G} \dot{H}+H^3 \dot{\G}\right)\right. \nonumber
\\
&& \left. -(k-1)(k-2) \dot{G}^2   \left(H^2-2 \dot{H}\right) +6 \G^2 \dot{H} \left(\dot{H}+2 H^2\right)\right] \ge 0    \nonumber \;\;\;\; \text{if} \;\; p<0
   \\
 DEC^{\;\;\; \nearrow}_{\;\;\; \searrow}&&\;\;\;\;\;\;\;\;\;\;\;\;\;\;\;\;\;\;\;\;\;\;\;\;\;\;\;\;\;\;\;\;\;\;\;\;\;\;\;\;\;{\text{and}} \nonumber
   \\
&&f_0 \G^{k-3} \left[-4 k (k-1)(k-2) \dot{\G}^2 \left(2 \dot{H}+5 H^2\right)-24 k \G^2 \left(6 H^2 \dot{H}+\dot{H}^2+7 H^4\right)  \right.   \label{EC2}
   \\
   && \left. -4 k (k-1) \G \left(2 \ddot{\G} \dot{H}+5 H^2 \ddot{\G} +4 H \dot{\G} \dot{H} +13 H^3 \dot{\G}\right) +\G^3\right]    \ge 0\nonumber \;\;\;\; \text{if} \;\; p\ge 0
\end{eqnarray}
\\
\begin{eqnarray}
&& 4 f_0 k \G^{k-3} \left[(k-1) \G \left(2 \ddot{\G} \dot{H} -H^2 \ddot{\G}+4 H \dot{\G} \dot{H}+H^3 \dot{\G}\right)\right. \nonumber
\\
&& \left. -(k-1)(k-2) \dot{G}^2   \left(H^2-2 \dot{H}\right) +6 \G^2 \dot{H} \left(\dot{H}+2 H^2\right)\right] \ge 0   \nonumber
   \\
 SEC^{\;\;\; \nearrow}_{\;\;\; \searrow}&&\;\;\;\;\;\;\;\;\;\;\;\;\;\;\;\;\;\;\;\;\;\;\;\;\;\;\;\;\;\;\;\;\;\;\;\;\;\;\;\;\;{\text{and}} \nonumber
   \\
&& f_0 \G^{k-3}\left[12 k (k-1)(k-2) \dot{\G}^2 \left(2 \dot{H}+H^2\right)+24 k \G^2 \left(10
H^2 \dot{H}+3 \dot{H}^2+7 H^4\right) \right.  \nonumber
\\
&& \left. +12 k (k-1) \G \left(2 \ddot{\G} \dot{H} +H^2  \ddot{\G}+4 H \dot{\G} \dot{H}+5 H^3 \dot{\G}\right)-\G^3 \right] \ge 0 \;.
    \label{EC3}
\end{eqnarray}
By plugging the cosmological expression of $\G$ and the field equations solution of $f(\G)$ cosmology into Eqs. \eqref{density} and \eqref{pressure}, it is possible to recast the effective energy density and the pressure in terms of the scale factor and $k$ only. Computations yield the plots in Fig. 1.
\begin{center}
\centering
\includegraphics[width=.99\textwidth]{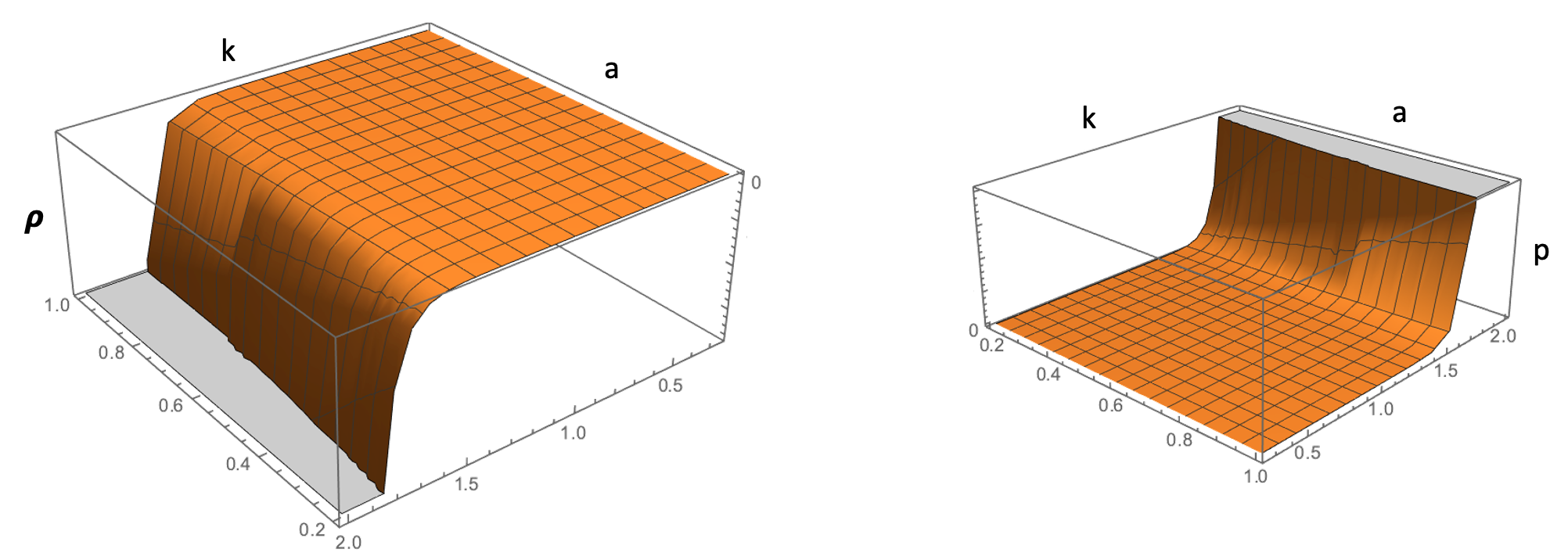}
Figure 1: \emph{Effective Energy Density (left panel) and Pressure (right panel) profiles as a function of $a$ and $k$.}
\end{center}
We set $f_0 = 1/2$ in order to recover the GR coupling. The effective EoS parameter can be obtained from the ratio $p/\rho$, which provides the expression
\begin{equation}
    w = \frac{1-12k}{3+12k},
\end{equation}
plotted in the graph of Fig. 2.
\begin{center}
\centering
\includegraphics[width=.59\textwidth]{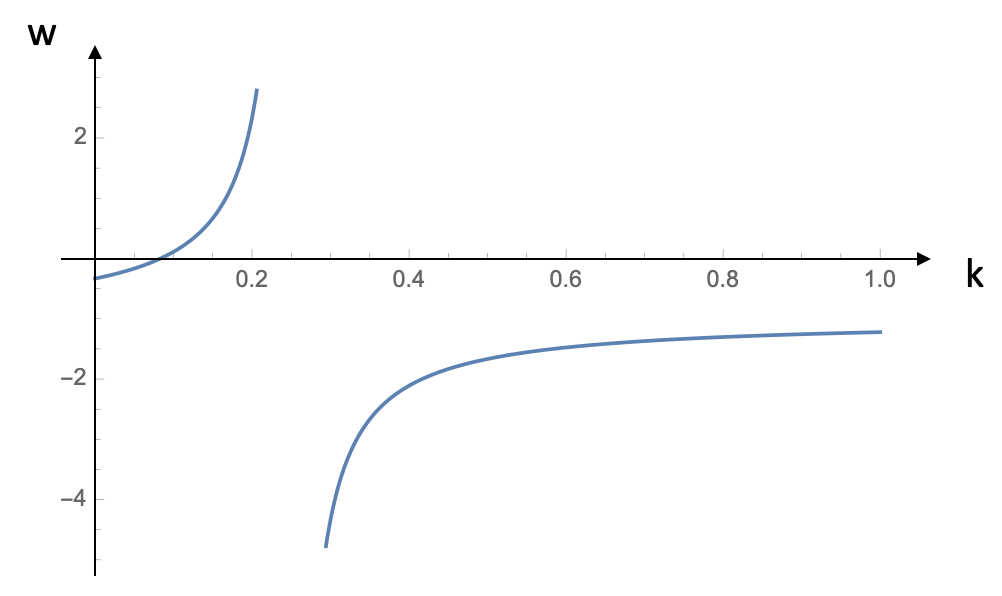}
\\ Figure 2: \emph{Effective EoS parameter in $f(\G)$ gravity}
\end{center}
Notice that the exact dark energy behavior is recovered only for large values of $k$. The effective EoS parameter is independent of the scale factor, as standard when considering time power-law solutions.
Now, with the aim to study the validity and the violation of the ECs and constrain the free parameter $k$, we introduce the cosmographic parameters $j$ (jerk), $q$ (deceleration) and $s$ (snap), defined as:
\begin{equation}
q = - 1 - \frac{\dot{H}}{H^2}   \;\;\;\; j = 1 + \frac{\ddot{H} + 3 \dot{H} H }{H^3} \;\;\;\; s =1 + \frac{\dddot{H} + 3 \dddot{H} H + 3 \dot{H}^2 + 6 H^2 \dot{H} + H \ddot{H}}{H^4} \,;
\label{jqs}
\end{equation}
in this way, the ECs can be recast in terms of $j, q, s, k$ only. To this purpose, notice that the cosmological expression of the Gauss-Bonnet scalar $\G = \displaystyle 24 H^2(H^2 + \dot{H})$ can be recast in terms of the cosmographic parameters as:
\begin{equation}
\G = -24 qH^4  \;\;\;\;\;\; \dot{\G} = 24 (3 q + 2 q^2 + j) H^5 \;\;\;\;\;\; \ddot{\G} = 24 (s - 12 q - 15q^2 - 2q^3 - 6j - 6jq) H^6 \; .
\label{Gparam}
\end{equation}
Therefore, the pressure and the energy density become:
\begin{eqnarray}
\rho_\G = \frac{f_0 2^{3 k-1} 3^k \left(-H^4 q\right)^k}{q^3} && \Big\{j^2 k
   (k-1)(k-2) +2 j (k-1) k q [k (2 q+3)-q-4] \nonumber
   \\
   && +q [k^3 q (2
   q+3)^2-k^2 (10 q^3+25 q^2+21 q+s)+k (6 q^3+9 q^2+15
   q+s)+q^2]\Big\}, \nonumber
\\ \nonumber \\
p_\G = \frac{f_0 2^{3 k-1} 3^{k-1} \left(-H^4 q\right)^k}{q^2} && \Big\{2 j^2 k
   (k-1)(k-2) +j k (k-1) [4(2 k-1) q^2+4 (3 k-4) q+3] +2 k^3 q^2 (2 q+3)^2 \nonumber
   \\
   && -k^2 q (20 q^3+50 q^2+36 q+2
   s-9)+2 k q (6 q^3+10 q^2+15 q+s-6)-3 q^2]\Big\},
\end{eqnarray}
respectively. Though the value of the term $\left(- q H^4 \right)^{k}$ depends on the specific $k$ considered, it is a total factor that multiplies all the rest of the inequality. Moreover, the numerical value of the deceleration parameter is negative, thus such total term can be neglected in the computation of the ECs. First we choose $f_0 > 0$, in order to recover the GR coupling as soon as $k = 1/2$. Considering Eqs. \eqref{jqs} and \eqref{Gparam}, and using the experimental values of $j,q,s$ provided in \cite{Rapetti:2006fv} ($q = -0.81$, $j = 2.16$, $s = -0.22$), the ECs are satisfied for:
\begin{equation}
\begin{split}
&\text{NEC} \to  k \le 0 \, \lor \,  1.457 \le k \le 2.977
\\
& \text{WEC} \to k \le 0
\\
&\text{DEC} \to k \le 0
\\
&\text{SEC} \to 1.457 \le k \le 2.977.
\end{split}
\end{equation}
It is worth remarking that the above result holds independently of the value of the Hubble constant $H$, since the latter multiplies all the ECs as a total factor with an even power. As we can see, no value of $k$ simultaneously satisfies all the ECs when $f_0 > 0$. Considering $f_0 < 0$, the ECs provide
\begin{equation}
\begin{split}
&\text{NEC} \to  0 \le k \le 1.457 \, \lor \,  k \ge 2.977
\\
& \text{WEC} \to 0.093 \le k \le 1.457 \, \lor \,  k \ge 2.977
\\
&\text{DEC} \to 0.113 \le k \le 1.457 \, \lor \,  k \ge 2.977
\\
&\text{SEC} \to 0 \le k \le 0.189,
\end{split}
\end{equation}
and all of them are simultaneously satisfied for $k$ comprehended in the strict range
\begin{equation}
0.113 < k < 0.189.
\label{strictrange}
\end{equation}
By considering the vacuum solution of the field equations in $f(\G)$ gravity \cite{Bajardi:2020osh}, we notice that also the scale factor is constrained by the ECs. Taking into account the result in Eq. \eqref{strictrange}, and the vacuum solution of $f(\G) \sim \G^k$ field equations, namely
\begin{equation}
a(t) = a_0 t^{1-4k},
\label{a(t)}
\end{equation}
the scale factors not violating the ECs are of the form:
\begin{equation}
a(t) = a_0 t^n \,\,\, \text{with} \,\,\, 0.24 \le n \le 0.55.
\end{equation}

Nevertheless, the ECs validity for other values of $k$ and $n$ might occur in a different epoch, where the behavior of the gravitational interaction was different than the current one. In this case, the function labeling the theory might assume the form
\begin{equation}
f(\G) = f_0 \G^{k_1} + f_1 \G^{k_2}.
\end{equation}
For weakly time-depending coupling constants, the possibility that at some epochs the contribution of $f_1$ ($f_0$) was predominant with respect to that of $f_0$ ($f_1$), can be taken into account. In summary, all the ECs are violated when $f_0 > 0$; on the other hand, when $f_0 < 0$ they are satisfied if $k$ lies within the range $0.113 < k < 0.189$, meaning that almost all values of $k$ (there including $k=1/2$) lead to a gravitational model mimicking the features of the Einstein theory with cosmological constant.
\subsection{Slow--Roll Inflation} \label{SRinfl}
In this subsection we use the previously written ECs, in order to find out the expression of the slow--roll parameters $\varepsilon$ and $\eta$ in terms of $k$. Specifically, we impose the magnitude of the slow--roll parameters to be small, namely consider the inequalities
\begin{equation}
|\varepsilon| =  \left|-\frac{\dot{H}}{H^2}\right| \ll 1 \quad |\eta| = \left|-\frac{\ddot{H}}{2 H \dot{H}} \right| \ll 1,
\label{SRCond}
\end{equation}
which have to be satisfied during inflation. The first condition comes from the requirement for an accelerated expansion of the universe through the relation 
\begin{equation}
  \frac{\ddot{a}}{a} = H^2 \left(1 + \frac{\dot{H}}{H^2} \right) = H^2 \left(1 - \varepsilon \right)> 0.
\end{equation}
The second condition arises from the minimal coupling between geometry and the scalar field. Specifically, starting from the action 
\begin{equation}
    S = \int \left[\frac{R}{2} + \frac{1}{2} \dot{\phi}^2 - V(\phi) \right] d^4 x,
    \label{NMA}
\end{equation}
in a cosmological spatially flat background the Klein--Gordon equation reads
\begin{equation}
\ddot{\phi} + 3 H \dot{\phi} + V_\phi(\phi) = 0,
\end{equation}
with $V_\phi(\phi)$ being the first derivative of $V(\phi)$ with respect to $\phi$. Inflation occurs when the potential energy is predominant with respect to the other terms. Then, the potential should have a minimum at the end of the inflation. In the slow--roll approximation, where the scalar field is expected to roll slowly, the field equations together with the Klein--Gordon equation yield the condition
\begin{equation}
- \frac{\ddot{\phi}}{H \dot{\phi}} = -\frac{\ddot{H}}{2 H \dot{H}} \ll 1.
\end{equation}
Similarly, in modified theories of gravity, the extra geometric terms can play the role of kinetic energy and potential of some time--depending scalar field.  

In order to impose the conditions \eqref{SRCond}, we rewrite the first two components of the field equations \eqref{field eq} as:
\begin{eqnarray}
 H^2 = \frac{1}{3 (2 R -4 \Box ) f_\G(\G)} && \left\{ \frac{f_0}{2} \G^{k-3} \left[-24 k (k-1) H^2 \left(\G \ddot{\G}+(k-2) \dot{\G}^2\right) \right. \right. \nonumber 
 \\
 && \left. \left. -72 k \G^2 H^4-48 k \G H^2 \left(2 \G (\dot{H} + H^2) +(k-1) H \dot{\G}\right)+\G^3 \right] \right\} ,
\end{eqnarray}

\begin{eqnarray}
\dot{H} =- \frac{3}{2} H^2 -\frac{1}{2(2 R -4 \Box ) f_\G(\G)} && \left\{\frac{f_0}{2} \G^{k-3} \left[16 k(k-1) (\dot{H} + H^2) \left(\G \ddot{\G}+(k-2) \dot{\G}^2\right) \right. \right. \nonumber
\\
&&  + 24 k \G^2 H^4+16 k \G (\dot{H} + H^2) \left(3 \G (\dot{H} + H^2)+2 (k-1) H \dot{\G} \right) \nonumber
\\
&&  \left. \left. +24 k \G H^2 \left(4 \G (\dot{H} + H^2) +(k-1) H \dot{G}\right)-\G^3\right] \right\}.
\end{eqnarray}
Replacing the vacuum solution of $f(\G)$ gravity \eqref{a(t)} in the above equations and considering the explicit expression of the Gauss--Bonnet term 
\begin{equation}
    \G = 24 H^2(H^2+\dot{H}),
\end{equation}
the magnitude of the slow--roll parameters turn out to be
\begin{equation}
    |\varepsilon| = |\eta| = \left| \frac{1}{1-4k}\right| .
\end{equation}
The conditions for the inflation
\begin{equation}
    |\varepsilon| \ll 1 \qquad |\eta| \ll 1,
\end{equation}
provide the constraint
\begin{equation}
    k \ll 0 \, \lor \, k \gg \frac{1}{2}.
    \label{slowrollk}
\end{equation}
This means that cosmological inflation in $f(\G)$ gravity occurs only when $k$ is strictly negative or when it is much larger than $1/2$. Interestingly notice that the value $k = 1/2$ is the limit in which $f(\G) = \G^k$ gravity behaves like Einstein GR in a cosmological spatially flat background. From the definition of the cosmographic parameters \eqref{jqs} we also notice that the deceleration parameter can be written in terms of $k$ as:
\begin{equation}
    q = -1 + \frac{1}{4k - 1} ,
\end{equation}
so that when $k\ll 0 \, \lor \, k \gg 1/2$, $q$ turns out to be negative, as we expect. The negative value of $q$ corresponds to an accelerated expansion of the universe, in agreement with the result provided in Eq. \eqref{slowrollk}. Slow--roll conditions are not sufficient to establish whether the theory is able to fit the cosmological data, and a further study is thus necessary. Nevertheless, it is a first step aimed at verifying whether the theory might be a good candidate for the inflationary model. 
\section{Energy Conditions in $R + f(\G)$ Cosmology} \label{ECCOS1}
With the aim to compare results coming from $f(\G)$ gravitational model with those related to $R+ f(\G)$ gravity, we evaluate the ECs in the latter theory, starting from the action \eqref{actionr+f} with function \eqref{symmetryfunc1}, \emph{i.e.}
\begin{equation}
S = \int \sqrt{-g} \left(\frac{R}{2} + f_0 \G^k\right) \, d^4x
\label{actionr+gk}
\end{equation}
and study the ECs by using the numerical values of cosmographic parameters. In this case, the field equations can be written as
\begin{eqnarray}
&&G_{\mu \nu} = \left[T_{\mu \nu} + \frac{f_0}{2} \G^k g_{\mu \nu} - k f_0 \left(2R R_{\mu \nu} - 4 R_{\mu p} R^p_{\,\,\,\nu} + 2 R_\mu^{\,\,\, p \sigma \tau} R_{\nu p \sigma \tau} - 4 R^{\alpha \beta} R_{\mu \alpha \nu \beta}\right) \G^{k-1} \right.\nonumber
\\
&& \left. + k f_0 \left(2R \nabla_\mu \nabla_\nu +4 G_{\mu \nu} \Box - 4 R_{\{ \nu}^p \nabla_{\mu \}}\nabla_p + 4 g_{\mu \nu} R^{p \sigma} \nabla_p \nabla_\sigma - 4 R_{\mu \alpha \nu \beta} \nabla^\alpha \nabla^\beta \right) \G^{k-1}\right] \;,
\end{eqnarray}
so that the RHS can be intended as an effective energy-momentum tensor which vanishes as soon as $f_0 = 0$. In a cosmological spatially flat spacetime, the two non-vanishing components of the field equations read
\begin{eqnarray}
G_{0}^0&=& \rho_0 + \frac{f_0}{2}\G^{k-2} \left\{24 k H^2 \left[(k-1) H \dot{\G}-\G (\dot{H} + H^2)\right] +1\right\} \label{rhogf}
\\ \nonumber
\\
G_{1}^1 &=& - \left\{ p_0 - \frac{f_0}{2} \G^{k-3} \left[8 H (k-1) k \left(2 \G \left(H^2+\dot{H}\right) \dot{\G}+H \left(\G \ddot{\G}+(k-2) \dot{\G}^2\right)\right)-24 H^2 k \G^2   \left(H^2+\dot{H}\right)+\G^3 \right] \right\}, \nonumber
\\
\label{pGF}
\end{eqnarray}
where $\rho_0$ and $p_0$ are the density and the pressure of matter, respectively. As before, the second terms in the RHS of the above equations can be intended as effective energy densities and pressures of the gravitational field, whose ratio takes the form
\begin{eqnarray}
    w &=& \frac{p_\G}{\rho_\G} = \left\{\frac{1}{3 H^2 \left(\dot{H}+H^2\right) \left[-k H \ddot{H} +2 (1-2 k) H^2 \dot{H}+(1-2 k) \dot{H}^2+H^4\right]}\right\} \nonumber
    \\
   & \times & \big\{ 6 k H^5 \ddot{H}+(8 k-9) H^6 \dot{H}+2 k (2 k-1) \dot{H}^4+H^4 \left[k \dddot{H}+\left(16 k^2-9\right) \dot{H}^2\right] \nonumber \\
   &&+ 4 k^2 H \dot{H}^2 \ddot{H} +2 k (4 k-1) H^3 \dot{H} \ddot{H} +H^2 \left[(k-2) k \ddot{H}^2+\left(16 k^2-2 k-3\right) \dot{H}^3+k \dddot{H} \dot{H}\right]-3 H^8 \Big\}
    \label{ratiow}
\end{eqnarray}
Eqs. \eqref{rhogf} and \eqref{pGF} can be written in terms of the cosmographic parameters \eqref{jqs} as:
\begin{equation} 
\rho_\G = \frac{2^{3k-1} 3^k f_0 (-H^4 q)^k}{q^2} (k-1) [j k -q^2 + k q (3 + 2 q)] \qquad \qquad \qquad \qquad \qquad \qquad \qquad \qquad \qquad
\end{equation}  
\begin{eqnarray}    
p_\G = \frac{2^{3k-1} 3^{k-1} f_0 (-H^4 q)^k}{q^3}(k-1) && [j^2 k (k-2)+ 2 j k q (-3 + 3k + 2 kq) +  3 q^3 + k^2 q^2 (3 + 2 q)^2 \nonumber
\\
&& - k q (6 q + 3 q^2 + 2 q^3 + s)],
\end{eqnarray}
where we used the form of $\G$ in Eq. \eqref{Gparam}. The ratio in Eq. \eqref{ratiow}, in terms of the cosmographic parameters, becomes:
\begin{equation}
w =  \frac{j^2 k(k-2 ) + 2 j k q [-3 + k (3 + 2 q)] + q [3 q^2 + k^2 q (3 + 2 q)^2 - k (6 q + 3 q^2 + 2 q^3 + s)]}{3 q [j k + q (-q + 3k + 2 k q)]}.
\end{equation}

We assume the density and the pressure of matter fields to satisfy the ECs separately. Using the same values of the cosmographic parameters considered in the previous section, when $f_0 > 0 $ the ECs are satisfied for: 
\begin{equation}
\begin{split}
&\text{NEC} \to k \le 0 \, \lor \,  1 \le k \le  4.371
\\
& \text{WEC} \to k \le 0 \, \lor \,  1 \le k \le  4.371
\\
&\text{DEC} \to -1.866 \le k \le 0 \, \lor \, 1.573 \le k \le 4.371
\\
&\text{SEC} \to k \le -0.313 \, \lor \,  1 \le k \le  3.129 \, .
\end{split}
\label{ECsR+Fg}
\end{equation}
The above inequalities admits a common solution, that is 
\begin{equation}
 -1.866 \le k \le  -0.313 \, \lor \,  1.573 \le k \le 3.129.
  \label{Range1}
\end{equation}
Similarly to the previous case, Eq. \eqref{Range1} can be used to constrain the scale factor which solves the field equations analytically, that is:
\begin{equation}
a(t) = a_0 \exp\left\{\left[\frac{24^k f_0 (1-k)}{3}\right]^{\frac{1}{2-4 k}} \, t\right\}.
\label{EXPSF}
\end{equation}
In the first range of Eq. \eqref{Range1}, namely $ -1.866 \le k \le  -0.313$, the exponent is positive and the scale factor describes an exponentially accelerated universe. In the second range, when $ 1.573 \le k \le 3.129$, the term in the square bracket turns out to be negative, leading to a bouncing cosmological model. 

Notice, however, that the solution \eqref{EXPSF} holds only when $k \neq 1/2$, otherwise we have
\begin{equation}
a(t)_{\pm} =a_0 t^\frac{4 f_0^2 + 3 \pm \sqrt{3} \sqrt{16 f_0^2+3}}{2 \left(3 -2 f_0^2\right)}.
\label{solpowerlaw}
\end{equation}
In this case, the scale factor $a(t)_+$ describes an accelerating universe when
\begin{equation}
- \sqrt{\frac{3}{2}} < f_0 < \sqrt{\frac{3}{2}},
\label{Rangef01}
\end{equation}
while in $a_-(t)$ the power of $t$ is always negative independently of the value of $f_0$. Since $k = 1/2$ violates all the inequalities in Eq. \eqref{ECsR+Fg}, when $f_0 >0$ and $ k =1/2$ a power-law universe acceleration occurs and the ECs are violated. 

Assuming a negative coupling constant, namely $f_0 < 0$, the ECs yield the constraints
\begin{equation}
\begin{split}
&\text{NEC} \to k \le 0 \, \lor \,  1 \le k \le  4.371
\\
& \text{WEC} \to 0.630 < k < 1
\\
&\text{DEC} \to \nexists  k \in \mathbb{R}
\\
&\text{SEC} \to 0 \le k \le 1 \, \lor \,  k \ge 4.371 \, 
\end{split}
\end{equation}
and cannot be simultaneously satisfied for any real value of $k$. However, it is worth noticing that all those $k$ included in the range $1 < k < 4.371$ violate the SEC, which means that the geometric contributions in $R + f(\G)$ gravity can act as a repulsive source of gravitational field. Moreover, similarly to the previous case, the effective EoS parameter can be written as a function of $k$, after replacing the cosmological vacuum solution of this theory into Eqs. \eqref{rhogf} and \eqref{pGF}. Computations yield
\begin{equation}
    w = -1 + \frac{4k}{3\ell}, \quad a(t) = a_0 t^\ell, \quad \ell = \frac{4 f_0^2+\sqrt{48 f_0^2+9}+3}{6-4 f_0^2}.
\end{equation}

\subsection{Slow-Roll Inflation}
For inflation to be realized in $R+f(\G)$ gravity, described by the action \eqref{actionr+gk}, we must evaluate the magnitude of the slow-roll parameters, as we did for the $f(\G)$ model. However, here we focus on a subcase of Eq. \eqref{actionr+gk}, setting $k = 1/2$. This is due to the fact that the action 
\begin{equation}
S = \int \sqrt{-g} \left( \frac{R}{2} + f_0 \sqrt{\G} \right) d^4x,
\label{actionsymm1}
\end{equation}
is the only one yielding time power-law solution for the scale factor, as pointed out in Eq. \eqref{solpowerlaw}. Moreover, within all possible actions of the form \eqref{actionr+gk}, the one in Eq. \eqref{actionsymm1} is the only containing Noether symmetries, according to the results of Sec. \ref{sec:noeth}. In a cosmological spatially flat background, the field equations can be written as:
\begin{eqnarray}
H^2 &=&\frac{f_0}{6}\G^{-\frac{3}{2}} \left\{12 H^2 \left[-\frac{1}{2} H \dot{\G}-\G (\dot{H} + H^2)\right] +1\right\} \label{FEINFL1}
\\ \nonumber
\\
\dot{H} &=&-\frac{3}{2} H^2 + \frac{f_0}{4} \G^{-\frac{5}{2}} \left[-12 H^2  \G^2   \left(H^2+\dot{H}\right)+\G^3 -2 H  \left(2 \G \left(H^2+\dot{H}\right) \dot{\G}  +H \left(\G \ddot{\G} -\frac{3}{2} \dot{\G}^2\right)\right) \right] . \nonumber
\\
\end{eqnarray}
With the aim to constrain the value of the coupling constant $f_0$ to those admitting slow-roll inflation, we consider the vacuum solution \eqref{solpowerlaw}, namely
\begin{equation}
a_{\pm}(t)=a_0 \, t^\frac{4 f_0^2 + 3 \pm \sqrt{3} \sqrt{16 f_0^2+3}}{2 \left(3 -2 f_0^2\right)}.
\end{equation}
Replacing $a_+$\footnote{No coupling constants are selected when evaluating $a_-(t)$.} in the field equations \eqref{FEINFL1} and considering the conditions for the inflation in Eq. \eqref{SRCond}, it turns out that slow-roll inflation is admitted by $R + f_0 \sqrt{\G}$ theory of gravity when  
\begin{equation}
f_0 \sim \pm \sqrt{\frac{3}{2}}
\label{Rangef0}
\end{equation}
In particular, the more $f_0$ approaches the values $\pm \sqrt{3/2}$, the faster the scalar field rolls down the potential hill. These values are in agreement with the range provided by the ECs in Eq. \eqref{Rangef01}. 
\section{Conclusions and Remarks} \label{concl}
We considered the ECs in two extended theories of gravity, both depending on the Gauss--Bonnet topological term. First, we did not impose the GR limit as a requirement, to subsequently include the scalar curvature into the starting action. In both cases, we considered a function selected by symmetry considerations, that is $f(\G) = f_0 \G^k$, and investigated the ECs violation in terms of the parameter $k$. Unlike GR, the ECs in modified theories of gravity assume non-trivial expressions, and for this reason the study of the ECs is useful for several reasons, when alternatives to GR are considered. In cosmology, they can suggest whether the extra geometric contributions can play the role of Dark Energy, though a careful choice of cosmographic parameters  $j,q,s$ is often important. We showed that when a function of the form $f(\G) = f_0 \G^k$ is considered as the only contribution into the action, the requirement for ECs validity selects a strict range of $k$. On the other hand, we also selected the interval of $k$ within which slow--roll inflation occurs. 

Similar results can be found also when the scalar curvature is added from the beginning. Specifically, it turns out that, when $ k = 1/2$, all the ECs are violated and power-law inflation occurs for a coupling constant $f_0$ approaching the values $f_0 \sim \pm \sqrt{3/2}$. Generally, studying the ECs in modified theories of gravity is useful to completely discard many theories and to show that the validity of the ECs is not as trivial as the GR case. In the late-time, modified theories of gravity can address the observed acceleration of the Universe as a curvature effect, which can be understood as a fluid with negative pressure playing the role of Dark Energy. For this reason, the ECs violation in modified theories of gravity does not have to be intended in the same way as GR. While in GR the ordinary matter is the only component in the RHS of the field equations, modified theories of gravity admit effective energy density and pressure of the gravitational field. As the ordinary matter must satisfy all the ECs, some of them can be violated by the effective energy-momentum tensor of the gravitational field. In future works we aim to extend the prescription to a general function of both $R$ and $\G$, with the aim to constrain the action by means of the ECs and investigate the slow--roll inflation in the early time. 

\section*{Acknowledgments}

The Author acknowledges the support of {\it Istituto Nazionale di Fisica Nucleare} (INFN) ({\it iniziativa specifica} GINGER

\end{document}